\documentclass{caosp302}

\usepackage{graphicx}

\articleNo{123}
\pubyear{2005}
\volume{35}
\volnumber{3}
\firstpage{1}
\received{May 1, 2005}
\accepted{August 28, 2005}

\begin{document}

%
\hauthor{J.\,Silvester {\it et al.}}

\title{Magnetic Doppler Imaging of Ap stars}


%
\author{
        J.\,Silvester \inst{1,} \inst{2,} 
      \and 
        G.A.\,Wade \inst{1}   
      \and 
        O.\,Kochukhov \inst{3}
       \and 
       J.D.\, Landstreet \inst{4}
       \and 
	S.\, Bagnulo \inst{5}
 }

%
\institute{
           Department of Physics, Royal Military College of Canada, PO Box 17000, Station "Forces", Kingston, Ontario, Canada K7K 4B4  \\ \email{james.silvester@rmc.ca}
         \and 
           Department of Physics, Queen's University, Kingston, Ontario, Canada\\
         \and 
           Uppsala Astronomical Observatory, Box 515, 751 20 Uppsala, Sweden\\
         \and
	   Department of Physics \& Astronomy, The University of Western Ontario, London, Ontario, Canada, N6A 3K7 \\
         \and
          Armagh Observatory, College Hill, Armagh BT61 9DG, Northern Ireland \\
    }

\date{March 8, 2003}

\maketitle

\begin{abstract}
Historically, the magnetic field geometries of the chemically peculiar Ap stars were modelled in the context of a simple dipole field. However, with the acquisition of increasingly sophisticated diagnostic data, it has become clear that the large-scale field topologies exhibit important departures from this simple model. Recently, new high-resolution circular and linear polarisation spectroscopy has even hinted at the presence of strong, small-scale field structures, which were completely unexpected based on earlier modelling. This project investigates the detailed structure of these strong fossil magnetic fields, in particular the large-scale field geometry, as well as small scale magnetic structures, by mapping the magnetic and chemical surface structure of a selected sample of Ap stars. These maps will be used to investigate the relationship between the local field vector and local surface chemistry, looking for the influence the field may have on the various chemical transport mechanisms (i.e., diffusion, convection and mass loss). This will lead to better constraints on the origin and evolution, as well as refining the magnetic field model for Ap stars. 
Mapping will be performed using high resolution and signal-to-noise ratio time-series of spectra in both circular and linear polarisation obtained using the new-generation ESPaDOnS and NARVAL spectropolarimeters at the CFHT and Pic du Midi Observatory. With these data we will perform tomographic inversion of Doppler-broadened Stokes IQUV Zeeman profiles of a large variety of spectral lines using the INVERS10 magnetic Doppler imaging code, simultaneously recovering the detailed surface maps of the vector magnetic field and chemical abundances.

\keywords{MDI - Ap Stars - Magnetic Topology}
\end{abstract}

%
\section{Introduction}
\label{intr}

Magnetic fields play a fundamental role in the physics of the atmospheres of a significant fraction of stars on the H-R diagram. The magnetic fields of early-type stars (the Ap and Bp stars) have quite different characteristics and probably a different origin, than those of late-type stars like the sun (e.g. Mestel 2003). In early-type stars, the large-scale surface magnetic field is static on timescales of at least many decades, and appears to be ``frozen'' into a rigidly rotating atmosphere. The magnetic field is globally organised, permeating the entire stellar surface, with a relatively high field strength (typically of a few hundreds up to a few tens of thousands of gauss). The presence of the magnetic field strongly influences energy and mass transport (e.g., diffusion, convection and weak stellar winds) within the atmosphere of a star, and results in the presence of strong chemical abundance nonuniformities in photospheric layers, although to what level the magnetic field plays a role with these processes is still to be constrained.

One of the earliest discussion of the magnetic field geometry of an Ap star was published by Stibbs (1950) in his {\it  oblique rotator model} (ORM), in which the magnetic field is assumed to be a centred, pure dipole, usually inclined to the axis of stellar rotation, and ``frozen'' into the star. Later Preston (1969) discovered that the field geometry of HD 215441 deviated from the centred dipole assumed by Stibbs (1950) in the ORM. Leroy et al. (1994, 1996) were one the first to systematically study Ap stars using linear polarisation measurements, constraining the transverse component of the magnetic field and providing an unambiguous view of the large-scale geometry of the field. They concluded that the magnetic geometry of many of the stars could not be modelled in the context of a pure dipole field and that many stars displayed small-scale departures in their magnetic fields from the assumed smooth underlying low-order multipole.

The very first measurements of rotationally-modulated Zeeman circular and linear polarisation resolved within stellar line profiles (all 4 Stokes parameters obtained using the MuSiCoS spectropolarimeter at Pic du Midi Observatory) were reported by Wade et al. (2000).   This data set was later used with the new Magnetic Doppler Imaging technique (MDI), described by Piskunov \& Kochukhov (2002) and Kochukhov \& Piskunov (2002), to construct the very first assumption-free, high resolution maps of the surface vector magnetic field of the Ap stars 53 Cam (Kochukhov et al. 2004) and $\alpha^2$ CVn (Kochukhov et al.).

The unique maps of 53 Cam and $\alpha^2$ CVn reveal that the magnetic topology departs significantly from the commonly-assumed low-order multipolar geometry and, that the abundances were not distributed uniformly but localised in complex structures. The data used in the 53 Cam and $\alpha^2$ CVn maps represented the best data set obtained from several years of MuSiCoS observations. However,  with this data set the Stokes $Q$ and $U$ signatures were only really detectable in 3 strong lines, with a S/N of 5 or less in a sample of 2 or 3 stars.  The low signal-to-noise ratio and resolving power achievable with the MuSiCoS instrument led to significant ambiguity in the field reconstruction. Consequently, only an extremely limited range of stellar properties (rotation, mass, temperature, magnetic field, etc.) which may influence the phenomena of interest could be studied using the MuSiCoS data.

\subsection{The project}
To improve on the maps of Kochukhov et al., and to address important questions about field structure, field origin and the surface chemical transport mechanisms, we have begun to acquire new Stokes IQUV observations of a small sample of well-studied magnetic Ap stars using the ESPaDOnS and NARVAL spectropolarimeters. These instruments are designed to overcome the limitations encountered with MuSiCoS, with improved resolution, throughput and wavelength coverage. The ESPaDOnS spectropolarimeter is installed at the CFHT (Mauna Kea, Hawaii) and the NARVAL spectropolarimeter at l'Observatoire du Pic du Midi (France). ESPaDOnS and NARVAL are in fact identical instruments, with NARVAL being a copy of ESPaDOnS. They have high spectral resolving power ($R = \lambda / \Delta \lambda = $ 65000), allowing the analysis of the high resolution spectrum at the same time as any of the 4 Stokes states of polarisation. 
 
With the vastly improved data set we will create a new generation of MDI maps (using INVERS10) with which we will be able to characterize the magnetic field geometry of Ap stars and further refine the current model of the magnetic field structure. In addition the maps will probe the abundance surface structure of these stars, allowing for connections between the magnetic field and structure formation processes (diffusion, convection, mass loss) to be drawn. Because of the increased sensitivity of NARVAL/ESPaDOnS compared to MuSiCoS, we can probe a larger sample than was ever possible before. 

In addition to constructing maps of the magnetic field and chemical abundance distributions, we also intend to use the new measurements to verify the polarimetric performance of ESPaDOnS and NARVAL, to check the quality and consistency of the lower-quality MuSiCoS maps, and to search for evolution of the magnetic field (as suggested by recent MHD simulations; Braithwaite \& Spruit 2004).

\subsubsection{Target Selection}
All of our targets are well-established Ap stars with strong magnetic fields, spanning a significant range of mass and rotation rate. This allows us to study the dependence of the magnetic field configuration and surface abundance distributions on these fundamental parameters. These stars have well-determined rotation periods and projected rotational velocities.

{\tiny
\begin{table}
\begin{tabular}{ccccccc}
\hline
\hline
HD    & $\alpha$ (2000) & $\delta$ (2000) & V & T$_{\rm eff}$  & Period & $v \sin i$  \\
      &                 &                  &  &                 & (Days)    & (km/s) \\
\hline
  4778 & 00:50:17.91 & +45:00:07.6 & 6.15 &  10000 & 2.56 & 42\\
 32633 & 05:06:08.24 & +33:55:08.6 & 7.07 &   12800 & 6.43 & 20  \\
 40312 ($\theta$ Aur)  & 05:59:43.27 & +37:12:45.3 & 2.62 &  10200 & 3.61 & 54   \\
 62140 (49 Cam)& 07:46:27.64 & +62:49:53.0 & 6.49 &    7700 & 4.29 & 23   \\
 65339 (53 Cam)& 08:01:42.45 & +60:19:27.7 & 6.01 &   8300 & 8.03 & 13\\
 71866 & 08:31:10.64 & +40:13:29.6 & 6.76 &   8800  & 6.80 &    14  \\
112413 ($\alpha^2$ CVn)& 12:56:02.59 & +38:19:03.4 & 2.90 &   11500 & 5.46 & 14 \\
118022 (78 Vir) & 13:34:07.93 & +03:39:32.3 & 4.94 &   9100  & 3.77 & 21 \\
\hline\hline
\end{tabular}
\caption{Target Ap stars to be observed for this project}
\label{stars}
\end{table}
}

The target S/N for our spectra is greater than 500:1, which, for typical Stokes $Q/U$ amplitudes of 2\%, provides a relative S/N of the profile of about 10:1 - the value that we judge as necessary for a significant improvement in the quality of MDI maps. Relatively dense rotational phase sampling is required for optimal MDI reconstruction of the stellar surface (typically 15 or more Stokes IQUV observations uniformly-distributed over the stellar rotational cycle).

\subsection{Early Results}
\begin{figure}
\begin{center}
   \includegraphics[width=0.70\textwidth]{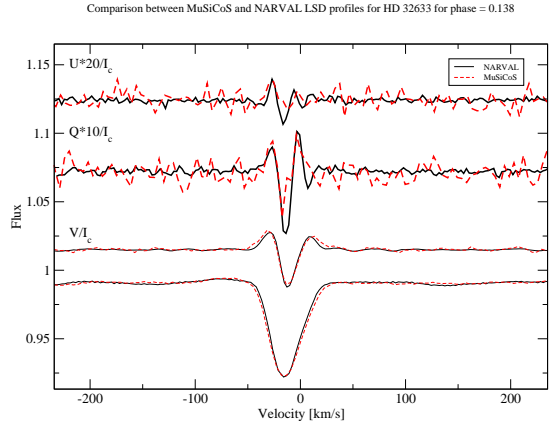}
   \caption{A comparison between Least-Squares Deconvolved (LSD) Stokes IQUV profiles for HD 32633 obtained with MuSiCoS and NARVAL, at rotation phase of 0.138. The MuSiCoS data is shown by a dashed line and the NARVAL data is shown as a solid line. The signatures are in good argeement. In Q and U the noise associated with the NARVAL data is clearly many times lower than that of the MuSiCoS data.}
\label{lsdc}
\end{center}
\end{figure}

\begin{figure}
\begin{center}
   \includegraphics[width=0.75\textwidth]{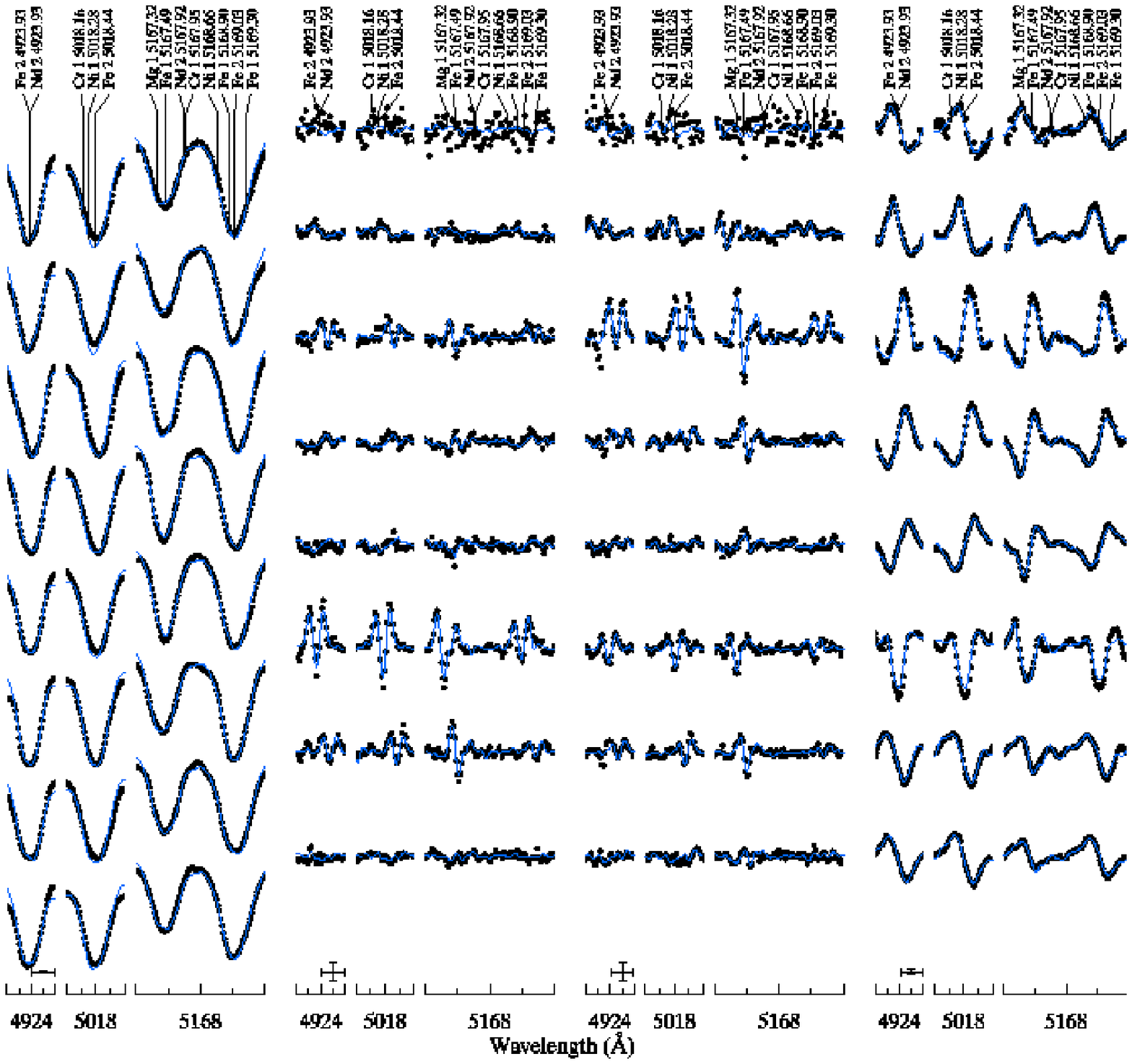}
   \caption{Preliminary magnetic field and intensity fit for 49 Cam, as recovered using MDI and the new IQUV spectropolarimetric data. The solid line is the model and dotted line is the observations, shown for 4 spectral lines as a function of rotational phase. }
\label{prof}
   \includegraphics[width=0.75\textwidth]{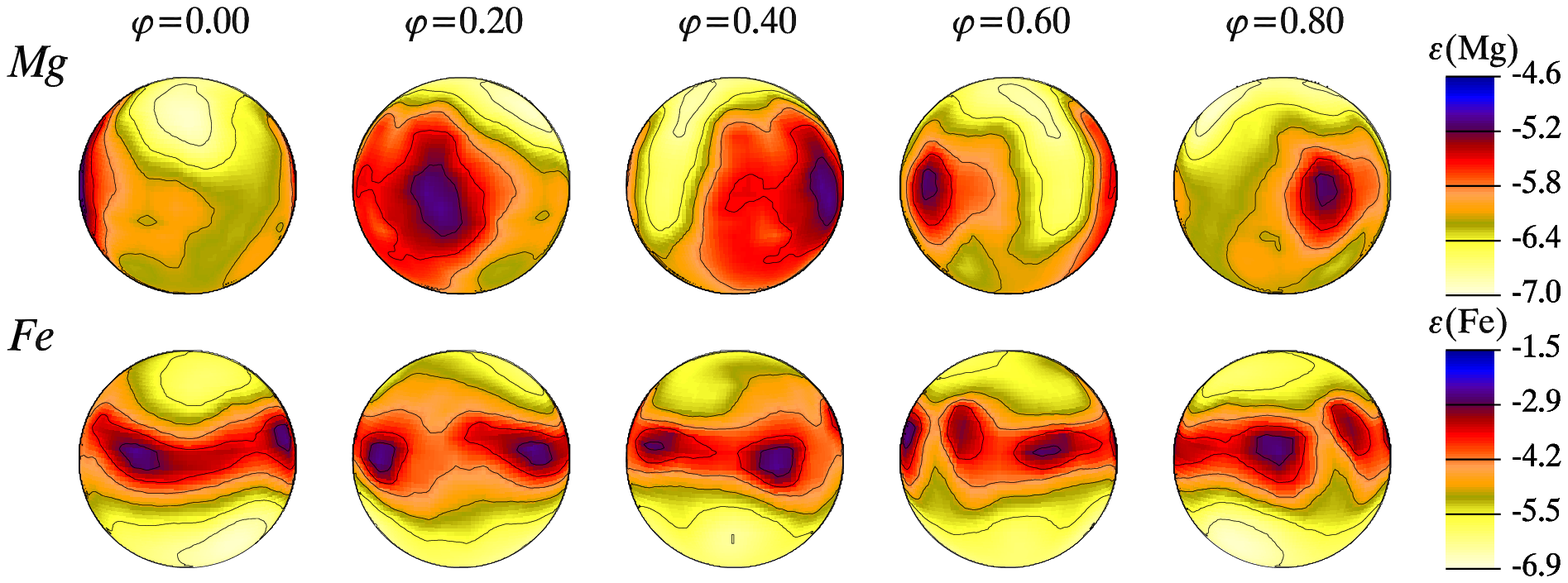}
   \caption{Abundance distributions of Mg and Fe on the surface of 49 Cam, as recovered simultaneously with the vector magnetic field using the MDI technique.}
\label{abu}
 \includegraphics[width=0.75\textwidth]{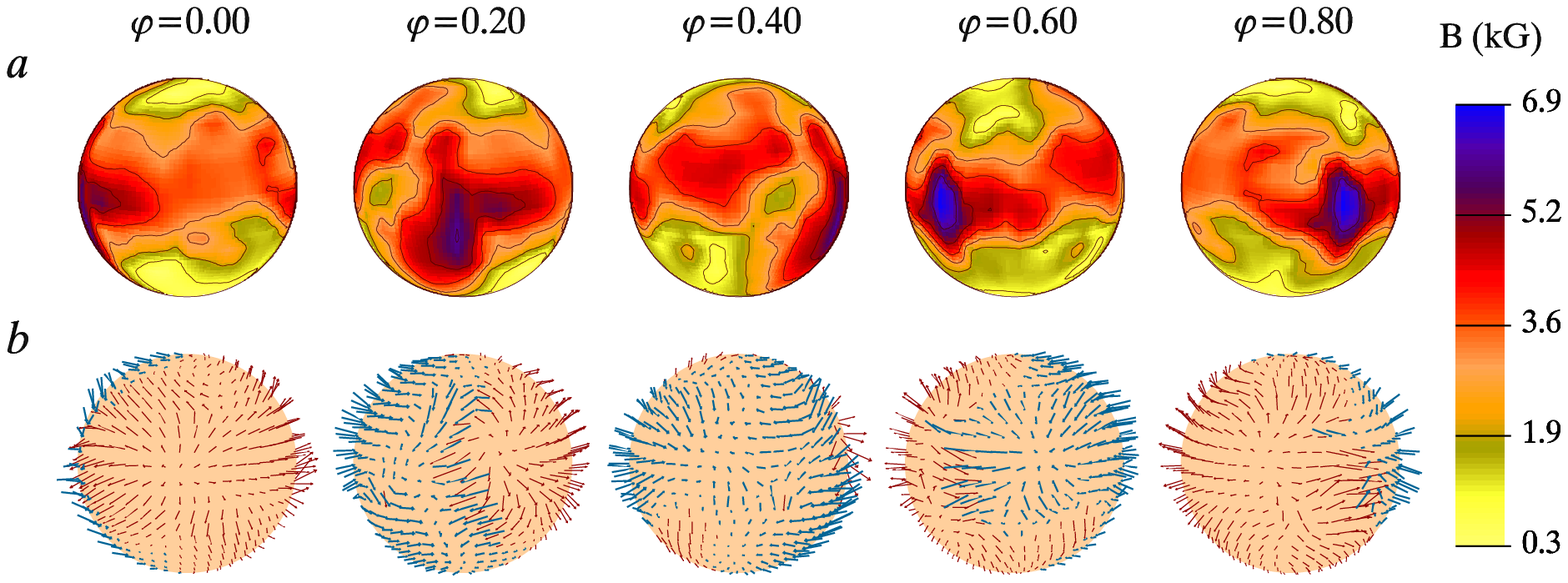}
   \caption{Preliminary magnetic field orientation and intensity maps of 49 Cam, as recovered using the
MDI technique and the new IQUV spectropolarimetric data.}
\label{fld}
\end{center}
\end{figure}

We have begun to measure the morphology and variation of the Zeeman circular and linear polarisation profiles within metallic spectral lines of our sample of magnetic Ap stars, with early results clearly demonstrating the improved quality of data (a direct comparison between the signal to noise obtained with MuSiCoS and NARVAL is illustrated in the Least-Squares Deconvolved Stokes profiles shown in Fig. \ref{lsdc}), with all observations exceeding the minimum signal-to-noise ratio of 500:1. In all the obtained spectra, clear variations in polarisation profiles are seen. 
With the first data obtained we were able to produce some early results and to test the MDI code with the new NARVAL and ESPaDOnS data. Early magnetic field and abundance maps have been produced for 49 Cam. These very preliminary maps are shown in Figs. \ref{fld} and \ref{abu}., with the fit between the model and observations shown in Fig. \ref{prof}.

\section{Conclusion}
With forty-two Stokes IQUV observations having already been obtained with ESPaDOnS and NARVAL, the instrument and data are proving to be of the excellent quality as predicted, and are overcoming the limitations of the MuSiCoS data set.
Observing time allocated for December 2007 with ESPaDOnS (4 nights) and Narval (8 nights) will allow for completion of the data sets for at least 3 of the stars.
Accurate stellar parameters still need to be determined for some of the stars (such as tilt and azimuth angle of the stellar rotational axis), for which reliable literature values are not available.
Within 6 months we will have the first in a new generation of MDI maps for a selection of new Ap stars, which will lead to refined models for the magnetic geometries these of stars, and will offer some clues to the true effect the field has on the abundance structures and the processes that form them.

\acknowledgements
JS would like to thank the Uppsala Astronomical Observatory for the support and technical assistance it has given to this project.



\end{document}